\documentclass[%
reprint,
superscriptaddress,
 amsmath,amssymb,
prb,
]{revtex4-2}

\usepackage{graphicx}
\usepackage{dcolumn}
\usepackage{bm}
\usepackage{pifont}
\usepackage[version=3]{mhchem} 
\usepackage{braket}
\usepackage{float}
\usepackage{xcolor}
\usepackage{mathtools}
\usepackage{tabularx}
\DeclareMathSymbol{*}{\mathbin}{symbols}{"03} 
\DeclareMathSymbol{\ast}{\mathbin}{symbols}{"03}
\usepackage{hyperref}
\hypersetup{
    colorlinks=true,
    linkcolor=blue,
    filecolor=blue,      
    citecolor=blue,
    urlcolor=blue,
    pdftitle={Overleaf Example},
    pdfpagemode=FullScreen,
    }

\begin{document}

\preprint{APS/123-QED}
\title{Unified Bonding Entropy  Model for Kekul\'{e} Graphene Nanoflakes}

\author{Chang-Chun He}
\email{scuthecc@scut.edu.cn}
\affiliation{School of Physics and Optoelectronics, South China University of Technology, Guangzhou 510640, China}
\author{Yu-Jun Zhao}
\affiliation{School of Physics and Optoelectronics, South China University of Technology, Guangzhou 510640, China}
\author{Xiao-Bao Yang}
\email{scxbyang@scut.edu.cn}
\affiliation{School of Physics and Optoelectronics, South China University of Technology, Guangzhou 510640, China}

\date{\today}

\begin{abstract}
The open-shell character of Kekul\'{e} graphene nanoflakes (GNFs) is conventionally 
rationalized by the gain of Clar aromatic $\pi$-sextets upon 
electron unpairing. While this rule successfully explains many 
quinoidal diradicaloids, it treats only the maximum number of 
sextets and neglects the multiplicity and spatial distribution of 
resonance configurations that realize the same Clar count. Here, 
we identify a second route to open-shell stabilization in which 
the maximum Clar-sextet number remains unchanged while the number 
of accessible Clar resonators increases substantially. We term 
this mechanism \emph{Clar-number-invariant resonance-space 
expansion}. By enumerating closed-shell and open-shell Clar 
resonators and combining this analysis with a bonding entropy 
model (BEM), we show that electron unpairing can release 
closed-shell pairing constraints, enlarge the resonance manifold, 
and redistribute C--C bond occupancies away from localized 
single- and double-bond limits. The BEM-predicted number and 
spatial distribution of unpaired electrons correlate strongly 
with density-functional-theory diradical character, local 
magnetic moments, optimized C--C bond lengths, and relative 
energies across a broad set of GNFs. The 
resulting framework offers a graph-based and physically 
transparent route for screening open-shell carbon 
nanostructures and for designing tunable molecular spins 
without requiring an increase in the maximum Clar number.
\end{abstract}

\maketitle

\section{Introduction}

While non-Kekul\'{e} GNFs such as triangulene and Clar's goblet have 
demonstrated intrinsic magnetism arising from sublattice imbalance 
or topological frustration~\cite{pavlicek2017synthesis,mishra2020topological}, 
their synthesis requires ultrahigh-vacuum on-surface techniques with 
limited scalability and ambient stability. In contrast, Kekul\'{e} 
diradicals retain a closed-shell resonance form and thus offer 
superior chemical stability and experimental 
accessibility~\cite{shu2023stable}. A recently synthesized 
tridecacyclic Kekul\'{e} diradicaloid achieves $y_0 = 0.98$, driven 
by a gain of five Clar's $\pi$-sextets, while retaining sufficient 
stability for experimental isolation~\cite{kuriakose2022design}. 
Many Kekul\'{e} diradicals also possess small singlet-triplet gaps 
that enable efficient triplet exciton harvesting, making them 
promising for thermally activated delayed fluorescence and molecular 
spintronics~\cite{noda2018excited,pan2024magnetooptical,%
weng2024biolympicenyl}. These practical advantages motivate a 
systematic understanding of the electronic origins of Kekul\'{e} 
radical formation.

Kekul\'{e} GNFs possess a perfect matching of the carbon lattice, so 
that all $\pi$ electrons can formally be paired, yet many such 
molecules still develop open-shell singlet or diradicaloid ground 
states. The conventional explanation invokes Clar's aromatic 
$\pi$-sextet rule as an empirical criterion: breaking a formal 
$\pi$ bond creates two radical centers, but the open-shell resonance 
form gains additional disjoint aromatic sextets whose stabilization 
is assumed to compensate for the energetic cost of 
unpairing~\cite{clar1972aromatic,bendikov2004oligoacenes,%
trinquier2018predicting}. This picture has qualitatively 
rationalized the diradical character of acenes and 
periacenes~\cite{yeh2016role,gopalakrishna2018open}, and synthesized 
Kekul\'{e} diradicals such as tetrabenzo[a,f,j,o]perylene indeed 
exhibit small singlet-triplet gaps~\cite{liu2015tetrabenzo,%
noda2018excited}. However, the Clar number compresses the full 
resonance manifold into a single integer, omitting the spatial 
distribution of sextets.  Graph-theoretical studies have established that the 
complete set of Kekul\'{e} and sextet patterns contains far more 
information than the Clar number alone~\cite{hosoya1981graph}. 
Consequently, two electronic states may share the same maximum 
Clar number yet possess fundamentally different resonance-space 
sizes and organizations. This raises a fundamental question that 
cannot be addressed by empirical sextet counting alone: 
\textbf{Can electron unpairing stabilize a Kekul\'{e} GNF by 
expanding the number of equal-Clar resonance configurations rather 
than by increasing the Clar number itself?} Answering this requires 
a quantitative model that captures the full statistical structure 
of the resonance manifold beyond the empirical Clar-counting picture.

To systematically investigate the magnetic properties of Kekul\'{e} 
GNFs, we combine explicit Clar-resonator 
enumeration~\cite{10.1021/acs.jpca.1c08661} with the bonding entropy 
model (BEM) developed in our recent 
work~\cite{he2025graph,he2025unified}. Within the BEM, all valence electrons are assigned to C--C bonds and unpaired-electron sites, and 
the optimal bond-occupancy numbers and spin-density distributions are 
obtained by maximizing a Shannon-type bonding entropy. This 
statistical procedure naturally compares the closed-shell and 
open-shell resonator ensembles and reveals two distinct routes to 
open-shell stabilization. The first is the Clar-sextet-gain 
mechanism, in which electron unpairing increases the maximum Clar 
number. The second is \textbf{Clar-number-invariant resonance-space 
expansion}: even when the maximum Clar number is unchanged, electron 
unpairing substantially increases the number of dominant resonators 
sharing the same sextet count by releasing closed-shell pairing 
constraints, thereby enabling Clar sextets to migrate among previously 
incompatible regions and delocalizing the C--C bond-occupancy 
numbers. If neither the Clar number nor the resonance space increases 
upon unpairing, the closed-shell state remains thermodynamically 
preferred. Crucially, the BEM furnishes a unified statistical 
framework that captures both routes through an identical 
entropy-maximization principle, without requiring \emph{a priori} 
knowledge of which mechanism is operative.

In this work,we validate this two-route criterion against density-functional-theory 
(DFT) calculations of diradical character $y_0$, local magnetic 
moments, optimized C--C bond lengths, and relative energies across a 
broad set of Kekul\'{e} GNFs. The BEM-predicted occupancy number and spatial 
distribution of unpaired electrons correlate strongly with the DFT 
observables, therefore provides a more complete criterion for 
Kekul\'{e} radical formation than the Clar number alone. The resulting 
framework offers a graph-based and physically transparent route for 
screening open-shell carbon nanostructures and for designing tunable 
molecular spins without requiring an increase in the maximum Clar 
number.

\section{Theoretical Framework and Computational Methods}

\subsection{Clar number and resonance-space size}
For a given molecular graph, we classify electronic configurations by 
the number of unpaired electrons $N_{u}$, where $N_{u}=0$ corresponds 
to the closed-shell state and $N_{u}>0$ to open-shell
states. For a fixed $N_{u}$, the Clar number 
$N_{\text{Clar}}(N_{u})$, defined as the number of mutually disjoint 
aromatic sextets, is uniquely determined by $N_{u}$. Because unpaired 
electrons release pairing constraints on the $\pi$ network, 
$N_{\text{Clar}}(N_{u})$ is generally a non-decreasing function of 
$N_{u}$, thereby providing a thermodynamic driving force for 
open-shell formation. The conventional Clar-sextet-gain descriptor is 
defined as the difference between the optimal open-shell and close-shell states:
\begin{equation}
\Delta N_{\text{Clar}}(N_{u})=N_{\text{Clar}}(N_{u})
-N_{\text{Clar}}(0),
\end{equation}
where $N_{u}^{\text{opt}}$ denotes the  number of unpaired 
electrons.

 For a fixed $N_{u}$ (and hence a fixed 
$N_{\text{Clar}}(N_{u})$), the same number of Clar sextets can be 
distributed over a GNF in many distinct spatial 
patterns. We therefore define $\Omega(N_{u})$ as the number of 
topologically distinct Clar resonators that attain the specific $N_{u}$:
\begin{equation}
\Omega(N_{u})=\bigl|\{C_{k}\,|\,N_{\text{sextet}}(C_{k};N_{u})
=N_{\text{Clar}}(N_{u})\}\bigr|,
\end{equation}
where symmetry-equivalent distributions $C_{k}$ are merged so that 
$\Omega(N_{u})$ counts distinct spatial arrangements of the sextets.


\subsection{Bonding entropy model}

The BEM assigns all valence electrons to C--C bonds, C--H bonds, and 
unpaired-electron sites, and determines the most probable distribution 
by maximizing a Shannon-type bonding 
entropy~\cite{he2025graph,PhysRevB.111.085408,10.1063/5.0280786}:
\begin{equation}
S=-\left[\sum_{i}^{N_{\text{bond}}}p_{i}\ln p_{i}
+\alpha\sum_{j}^{N_{C}}q_{j}\ln q_{j}\right],
\label{eq:bem}
\end{equation}
where $p_{i}=n_{i}/N_{\text{ele}}$ is the normalized occupancy number of 
bond $i$, $q_{j}=b_{j}/N_{\text{ele}}$ is the normalized 
unpaired-electron occupation on carbon site $j$, and $\alpha$ controls 
the statistical weight of the unpaired-electron sector~\cite{he2025unified}. The optimized 
C--C bond occupancy number $n_{ij}$ (with $1\le n_{ij}\le 2$) serves as a 
bond-order-like quantity. 

For every GNF, the bonding entropy was optimized as a function of the allowed 
number and spatial distribution of unpaired electrons, subject to the 
valence-electron constraint
\begin{equation}
\frac{n_{1}}{2}+\frac{n_{2}}{2}+\frac{n_{3}}{2}+b_{j}=4
\end{equation}
at each carbon atom, where $n_{1},n_{2},n_{3}$ are the electron counts 
in the three nearest-neighbor C--C or C--H bonds and $b_{j}$ is the 
number of unpaired electrons on that atom. 

\subsection{Resonator enumeration and electronic-structure calculations}

Clar resonators~\cite{10.1021/acs.jpca.1c08661} were generated by an 
in-house graph-theoretical algorithm that exhaustively searches for 
sets of mutually disjoint hexagonal rings satisfying the Clar rules, 
separately for the closed-shell and open-shell constraints. Radical 
sites were allowed at carbon vertices that preserve the 
$\pi$-conjugated framework, and symmetry-equivalent resonators were 
merged to avoid overcounting. The resulting dataset spans Kekul\'{e} 
GNFs with $\Delta N_{\text{Clar}}=0$--$3$ and multiple edge 
topologies. The implementation of the Clar-resonator 
enumeration and bonding entropy model is publicly available at 
GitHub~\cite{github_bem}.

Electronic-structure calculations were performed with Gaussian~16~\cite{g16}. Geometry optimizations and single-point 
energy evaluations employed the B3LYP exchange-correlation functional 
with the 6-31G(d) basis set~\cite{ditchfield1971self,hariharan1973influence}. All geometries were converged to tight 
criteria with ultrafine integration grids. Open-shell singlet states were treated using the spin-unrestricted 
broken-symmetry approach with a mixed initial guess to localize the 
two unpaired electrons on distinct subregions of the molecular graph. 
Spin contamination was monitored via $\langle S^{2}\rangle$ and 
corrected by approximate spin projection where appropriate. The 
diradical character $y_{0}$ was evaluated from natural-orbital 
occupations according to the Yamaguchi approximate spin-projection 
formula,
\begin{equation}
    y_{0}=1-\frac{2T}{1+T^{2}},\qquad 
    T=\frac{n_{\text{HONO}}-n_{\text{LUNO}}}{2},
\end{equation}
where $n_{\text{HONO}}$ and $n_{\text{LUNO}}$ are the occupation 
numbers of the highest occupied and lowest unoccupied natural 
orbitals, respectively. Local magnetic moments were obtained by 
Hirshfeld population analysis. Additional calculations for 
representative $\Delta N_{\text{Clar}}=0$ structures include 
singlet-triplet energy differences and benchmark tests with the 
$\omega$B97X-D range-separated functional to assess functional 
sensitivity.

\section{RESULTS and DISCUSSION}

\begin{figure*}
    \centering
    \includegraphics[width=\linewidth]{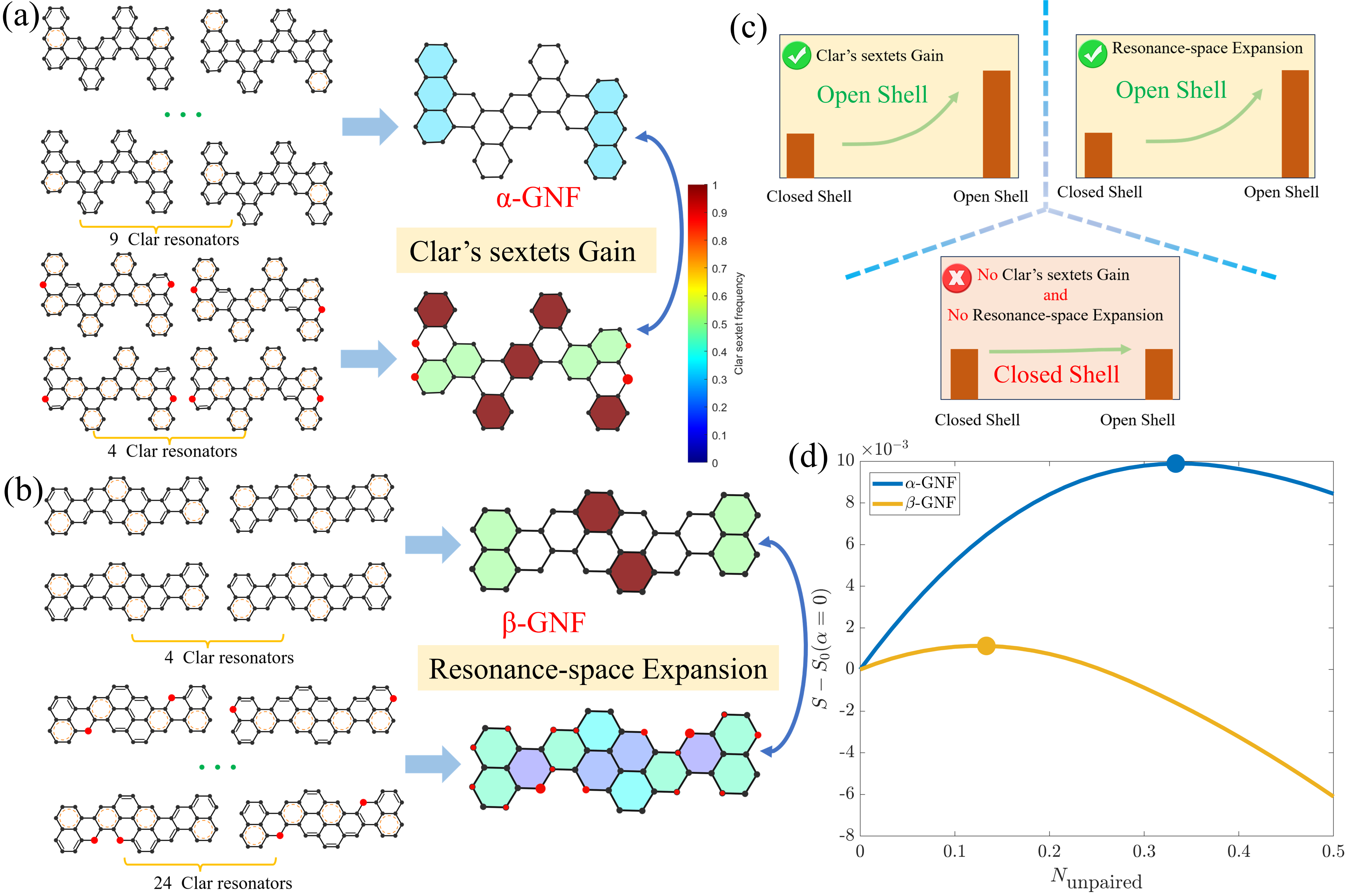}
\caption{\label{fig1} Two routes to open-shell stabilization in 
Kekul\'{e} GNFs. (a) Route~I (Clar-sextet gain): the tridecacyclic 
$\alpha$-GNF~\cite{kuriakose2022design} gains five Clar sextets 
upon electron unpairing ($N_{u}=2$). The sextet-occupancy map 
(right) shows enhanced local aromaticity relative to the 
closed-shell  state. (b) Route~II (resonance-space expansion): 
$\beta$-GNF retains the same maximum Clar number, but the dominant 
resonator count increases from 4 (closed-shell) to 24 (open-shell). The 
occupancy map (right) reveals that unpairing enables sextet 
coverage of all rings in the open-shell ensemble. (c) Two-coordinate 
criterion: a Kekul\'{e} radical forms when either Clar-sextet gain 
or resonance-space expansion is operative. (d) Relative bonding 
entropy $S-S_0$ versus $N_{u}$ at $\alpha=0$, where $S$ and $S_0$ 
are the bonding entropies of the open-shell and closed-shell 
states, respectively. The lower maximum for $\beta$-GNF indicates 
that resonance-space expansion provides weaker stabilization than 
Clar-sextet gain.}
\end{figure*}

Fig.~\ref{fig1} contrasts the two proposed routes to open-shell 
stabilization using representative molecular graphs.  For Route~I, we take the tridecacyclic polyaromatic hydrocarbon 
($\alpha$-GNF) recently synthesized by Kuriakose \textit{et 
al.}~\cite{kuriakose2022design} as an example. As shown in the left panel of Fig.~\ref{fig1}(a),  the 
molecule admits nine dominant Clar resonators in the closed-shell manifold, however, each of these 
configurations contains only double aromatic sextets, and 
the $\pi$ network remains globally under-aromatized. Upon electron 
unpairing, although the number of dominant resonators decreases to four,  
the sextet content of each structure increases by five. This gain of 
five Clar's $\pi$-sextets in the open-shell form provides sufficient 
aromatic stabilization to overcome the electron-pairing energy, 
leading to a high degree of diradical character ($y_{0}=0.98$). The 
energetic driving force is therefore primarily the increase in local 
aromaticity.

Additionally, to visualize the spatial distribution of aromatic stabilization, we 
overlay all dominant Clar resonators with equal statistical weight. 
For each hexagonal ring, we compute the probability that it hosts a 
Clar sextet across the resonator ensemble, and the  value of 1 indicates 
that the ring is aromatic in every resonator as depicted by the five hexagonal rings with deep red color, whereas lower values 
reflect fluctuating or absent sextet character. As shown in the right 
panel of Fig.~\ref{fig1}(a), the open-shell state with two unpaired 
electrons exhibits markedly higher local aromaticity than the 
closed-shell state. This quantitative 
difference in the sextet-occupancy map confirms that the energetic 
driving force for diradical formation in $\alpha$-GNF originates 
primarily from the gain in local aromaticity.

For Route~II, we construct a $\beta$-GNF structure in which the 
closed-shell and open-shell states share the same Clar number 
($\Delta N_{\text{Clar}}=0$). From the perspective of sextet count 
alone, the local aromatic content per resonator does not increase, 
as shown in the left panel of Fig.~\ref{fig1}(b). Nevertheless, 
this $\beta$-GNF develops genuine unpaired electrons. Electron 
unpairing expands the number of dominant Clar resonators from four 
in the closed-shell state to twenty-four in the open-shell state.  The 
maximum local aromatic content per resonator is unchanged, yet the 
global resonance manifold is dramatically enlarged.

This distinction is most clearly revealed by the sextet-occupancy 
map. As shown in the right panel of Fig.~\ref{fig1}(b), only six 
hexagonal rings are covered by Clar sextets in the closed-shell 
ensemble, whereas all rings acquire  sextet 
probability in the open-shell ensemble. The resonance-space 
expansion therefore homogenizes aromatic stabilization across the 
entire $\pi$ network, compensating for the electron-pairing 
penalty even though the Clar number remains invariant. 
This demonstrates that Clar-number-invariant resonance-space 
expansion alone can drive the emergence of unpaired electrons.

Fig.~\ref{fig1}(c) summarizes the proposed two-coordinate criterion for open-shell formation in Kekul\'{e} GNFs. A stable open-shell state can emerge via two distinct mechanisms: (i) a gain in the Clar number (Route I), or (ii) an expansion of the accessible resonance manifold at fixed Clar number (Route II). If neither the Clar number nor the resonance space increases upon electron unpairing, the closed-shell state remains thermodynamically preferred, because the electron-pairing penalty is not compensated by aromatic stabilization. 

To quantitatively capture the aromatic stabilization arising from 
resonance-space expansion, we employ the bonding entropy model (BEM) 
to describe the magnetic behavior of Kekul\'{e} GNFs. Within this 
framework, the optimal electron distribution is obtained by 
maximizing the bonding entropy $S$ at a fixed value of the weighting 
parameter $\alpha$, allowing us to assess whether the system favors 
a finite $N_{u}$. Remarkably, as shown in 
Fig.~\ref{fig1}(d), the relative bonding entropy $S-S_0$ for both 
$\alpha$- and $\beta$-GNFs first increases with $N_{u}$ even when 
$\alpha=0$ (i.e., unpaired electrons contribute nothing to $S$), 
where $S_0$ denotes the closed-shell bonding entropy. This initial 
rise reflects the gain in configurational entropy of the bonding 
electrons: electron unpairing releases pairing constraints and enables the 
$\pi$ electrons to access a broader resonance manifold, thereby 
delocalizing the bond-occupancy distribution and increasing the 
Shannon entropy of the C--C bonds. The subsequent decline at larger 
$N_{u}$ arises because the total number of electrons available for 
bonding decreases as more electrons localize as unpaired spins with $\alpha=0$. 
These localized spins contribute no entropy, so the 
loss of bonding electrons eventually dominates and $S$ decreases. 
The existence of a maximum at intermediate $N_{u}$ therefore 
provides a purely entropic criterion for open-shell formation, 
requiring that the gain in either Clar-sextet content or 
resonance-space size be sufficiently large to compensate for the 
reduction in bonding-electron entropy. This demonstrates that the 
BEM furnishes a unified quantitative description of both routes to 
diradical character, whether driven by Clar-sextet gain 
 or by Clar-number-invariant 
resonance-space expansion.

Notably, the entropy maximum of $\beta$-GNF in Fig.~\ref{fig1}(d) is 
lower than that of $\alpha$-GNF, indicating that the stabilization 
gained from Clar-number-invariant resonance-space expansion is 
weaker than that from Clar-sextet gain. This is consistent with 
the physical intuition that the migration of a fixed number of 
sextets among more configurations (Route~II) provides less 
energetic compensation than the outright increase in local 
aromatic content (Route~I). The BEM thus not only unifies the 
two routes qualitatively but also ranks their relative driving 
forces quantitatively.

\begin{figure}
    \centering
    \includegraphics[width=\linewidth]{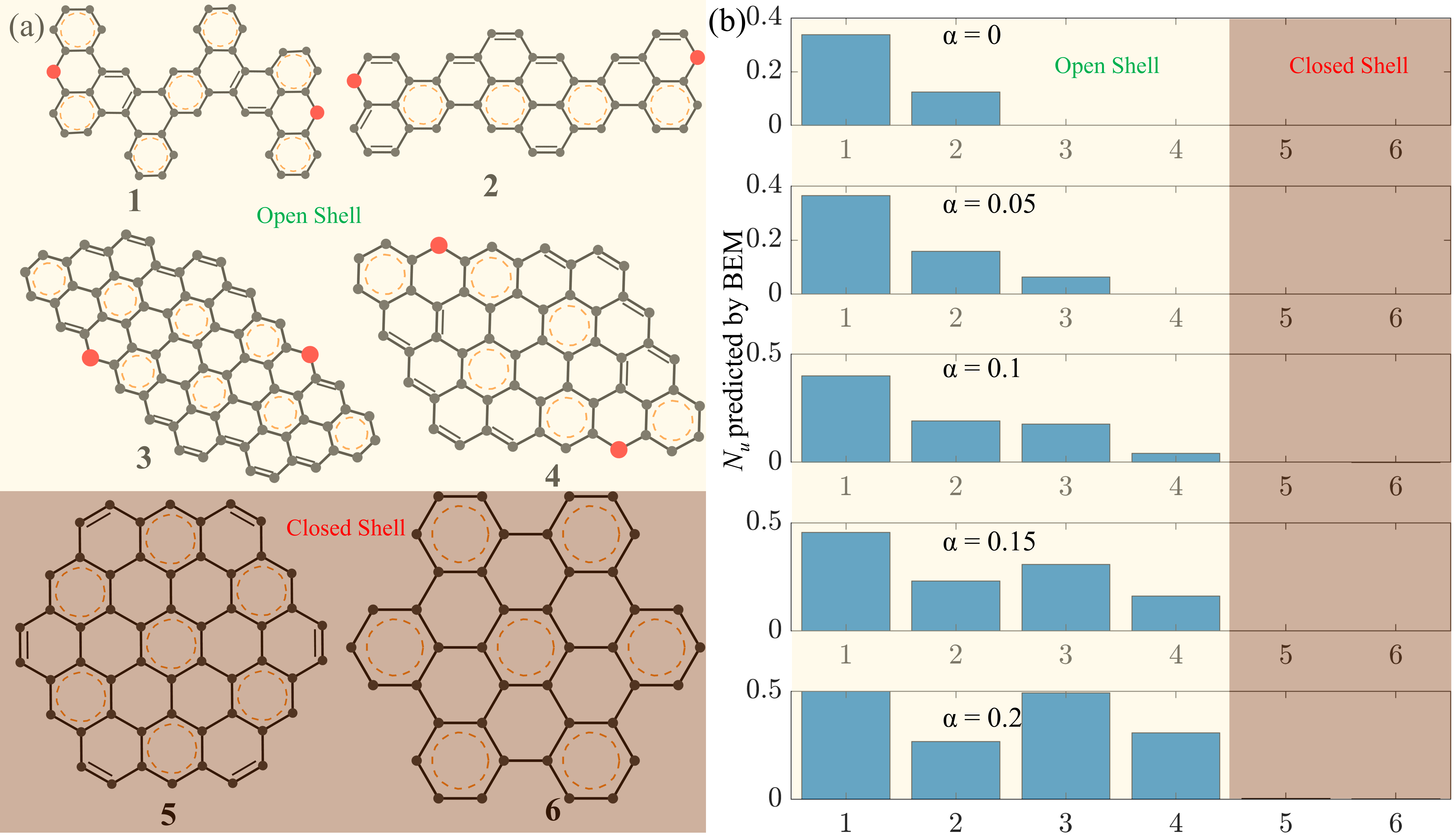}
\caption{\label{fig2}(a) Six representative Kekul\'{e} GNFs. Clar sextets 
(pink dashed circles) and unpaired-electron sites (red circles) are marked. \textbf{1} and \textbf{2} ($\alpha$-GNF and $\beta$-GNF) are robust open-shell; \textbf{3} and \textbf{4} are marginally open-shell; \textbf{5} and \textbf{6} are closed shell. (b) BEM-predicted $N_u$ versus 
$\alpha$ for \textbf{1} -- \textbf{6} GNFs.}
\end{figure}

It is important to emphasize that $\alpha$ is not an arbitrary 
fitting parameter,  yet it quantifies the relative statistical 
weight between unpaired electrons and bonding electrons in the 
entropy functional. To determine a physically reasonable value of 
$\alpha$ that enables the BEM to reliably distinguish open-shell 
from closed-shell Kekul\'{e} GNFs, we examine six representative 
structures spanning both magnetic and non-magnetic ground states, as 
shown in Fig.~\ref{fig2}(a). Because Kekul\'{e} GNFs possess a 
perfect matching, the $\alpha=0$ limit imposes a closed-shell 
pairing constraint that may artificially suppress the emergence 
of unpaired electrons even when the true ground state is open 
shell. A finite $\alpha$ is therefore required to allow the 
entropy gain from resonance-space expansion or Clar-sextet gain 
to overcome the electron-pairing penalty.

Structures \textbf{1} and \textbf{2} correspond to the $\alpha$-GNF and $\beta$-GNF discussed in Fig.~\ref{fig1}. As shown in Fig.~\ref{fig2}(b), the BEM at $\alpha=0$ already predicts finite unpaired-electron 
populations of $0.338$ and $0.125$ for these two structures, 
respectively, consistent with the entropy maxima at finite $N_{u}$ 
in Fig.~\ref{fig1}(d). This indicates that their open-shell 
character is sufficiently robust to be detected even with
zero weight of the unpaired-electron sector.

Structures \textbf{3} and \textbf{4}, by contrast, are confirmed by experiment~\cite{10.1021/jacs.2c11431} to possess 
open-shell ground states, yet the BEM at $\alpha=0$ predicts no 
unpaired electrons. A finite $\alpha$ is therefore necessary to 
recover their magnetic character. At $\alpha=0.05$, structure \textbf{3}
develops a finite $N_{u}$ whereas structure \textbf{4} remains closed shell. 
At $\alpha=0.1$, both structures exhibit non-zero unpaired-electron 
populations. Thus, an intermediate $\alpha$ in the range $0.1$--$0.2$ 
suffices to correctly classify structures \textbf{1}--\textbf{4} as open-shell state.

Structures \textbf{5} and \textbf{6} are canonical closed-shell GNFs. Remarkably, the 
BEM predicts $N_{u}=0$ across the entire scanned range 
$\alpha=0$--$0.2$, indicating that no amount of unpaired-electron 
weighting within this interval can induce a spurious open-shell 
solution. This robust closed-shell behavior arises because neither 
Clar-sextet gain nor resonance-space expansion is operative in 
these systems, and the pairing penalty is never compensated by 
aromatic stabilization.

Collectively, these results establish that $\alpha$ serves as a 
discriminating parameter that controls the balance between 
electron-pairing cost and resonance-driven entropy gain. A too small value 
($\alpha\leq 0.05$) fails to capture marginally 
stable radicals such as structure \textbf{4}, whereas a moderate value 
($\alpha\approx 0.1$--$0.2$) correctly separates the six 
structures into their respective magnetic classes. Based on this 
systematic scan and subsequent clustering analysis over a broader 
GNFs, we can determine the optimal $\alpha$ as the working value for the 
large-scale comparison in the following sections.

To systematically validate the predictive power of the BEM across a 
broad chemical space, we enumerated all Kekul\'{e} GNFs containing up 
to twelve fused hexagonal rings and evaluated their diradical character 
$y_{0}$ from unrestricted broken-symmetry DFT calculations. A 
structure was classified as possessing significant open-shell character 
when $y_{0}>0.1$. Within the BEM framework, we adopted $\alpha=0.15$ 
and used the threshold $N_{u}>0.1$ as the criterion for predicting 
unpaired electrons, and the calibration of this threshold is detailed in 
the Supplemental Material (SM).

\begin{figure}
    \centering
    \includegraphics[width=\linewidth]{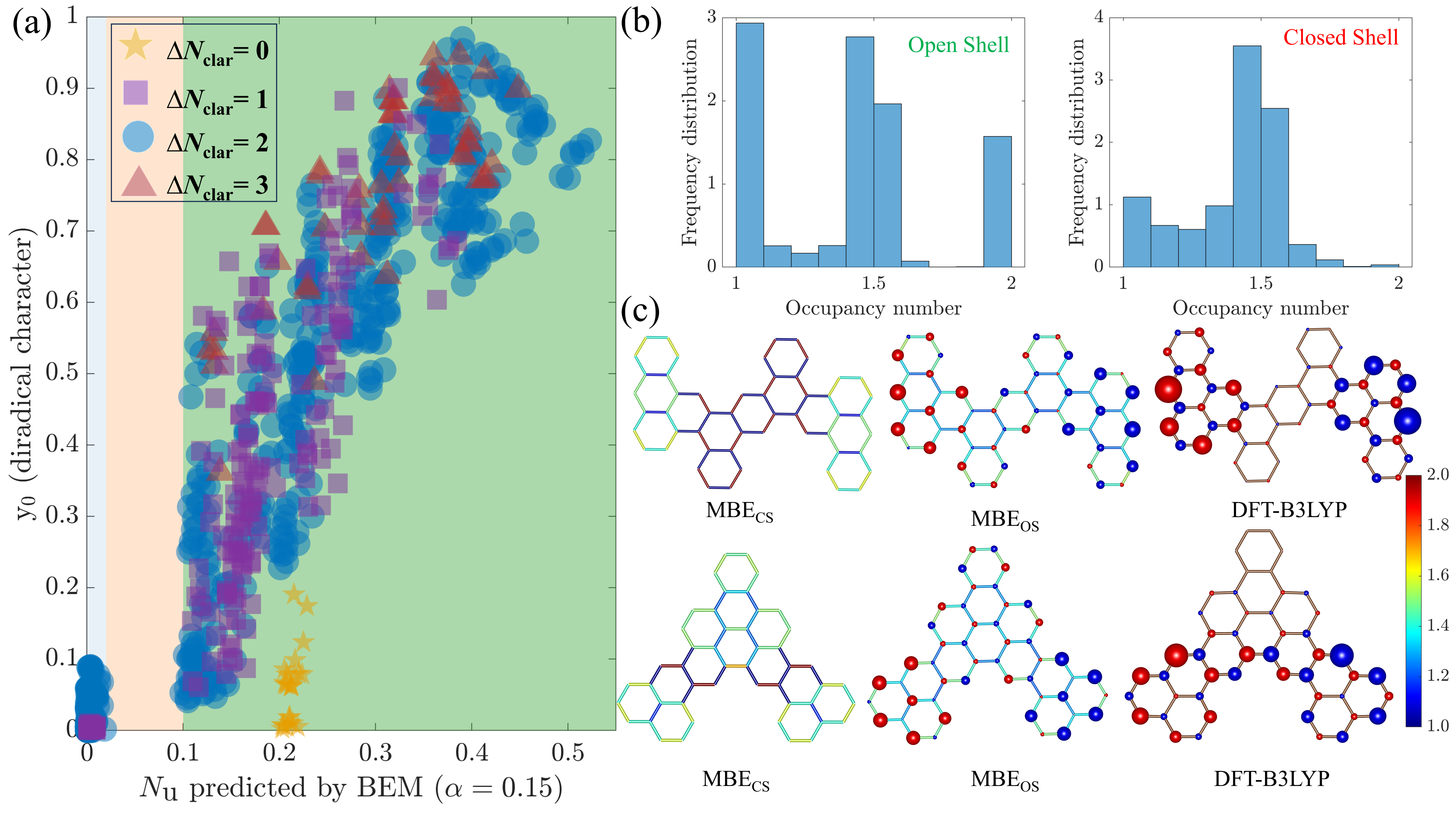}
\caption{\label{fig3}(a) DFT diradical character $y_{0}$ versus 
BEM-predicted $N_{u}$ ($\alpha=0.15$) for enumerated Kekul\'{e} GNFs.  (b) C--C bond 
occupancy number distributions under the closed-shell BEM constraint for open-shell structures  and closed-shell structures. 
(c) Two open-shell structures under closed-shell (left), 
open-shell (center) BEM and DFT-B3LYP solutions. C--C bond occupancy number are encoded 
by the color scale, and local magnetic moments are represented by spheres 
with radii proportional to their magnitude.}
\end{figure}

To quantify the predictive performance of the BEM, we analyze the 
correlation between the BEM-predicted $N_{u}$ ($\alpha=0.15$) and 
the diradical character $y_{0}$ for the full enumerated dataset. 
The Pearson correlation coefficient is $r=0.82$ and the Spearman 
rank correlation is $\rho=0.85$, indicating a strong monotonic 
relationship in Fig.~\ref{fig3}(a). The classification accuracy for 
distinguishing open-shell ($y_{0}>0.1$) from closed-shell 
($y_{0}\le 0.1$) GNFs is $94\%$ at the threshold $N_{u}=0.1$. 

Decomposing the data by the Clar-sextet gain $\Delta N_{\text{Clar}}$ reveals important 
differences in predictive variance. For $\Delta N_{\text{Clar}}=0$ 
structures, the correlation is weaker but remains significant as depicted by the yellow pentagons in Fig.~\ref{fig3}(a). Crucially, this BEM-based criterion 
enables the identification of magnetic GNFs that would be entirely 
missed by the conventional Clar-sextet-gain rule, for these structures 
possess genuine open-shell character yet exhibit 
$\Delta N_{\text{Clar}}=0$, rendering them invisible to any 
screening protocol that requires $\Delta N_{\text{Clar}}>0$ as a 
necessary condition. For $\Delta N_{\text{Clar}}\ge 0$, the 
correlation strengthens, consistent with the stronger energetic driving force provided by additional 
aromatic sextets.

As shown in Fig.~\ref{fig3}(a), the BEM-predicted $N_{u}$ 
($\alpha=0.15$) exhibits a clear positive correlation with the DFT 
diradical character $y_{0}$ across the enumerated GNFs. Notably, 
the scatter reveals significant limitations of the conventional 
Clar-sextet-gain rule. Even among structures with $\Delta 
N_{\text{Clar}}=1$ or $2$, many remain effectively closed shell 
($y_{0}\leq 0.1$), demonstrating that a mere increase in the 
maximum Clar number does not guarantee the emergence of unpaired 
electrons. Furthermore, structures sharing the same $\Delta 
N_{\text{Clar}}$ display a wide spread in $y_{0}$, indicating that 
the Clar-sextet count alone is insufficient to rank radical 
stability. The BEM captures this variance,  where the predicted $N_{u}$ distributes over a broad 
range for a given $\Delta  N_{\text{Clar}}$,, thereby providing a finer discrimination than the integer 
Clar-number difference.

More strikingly, the BEM successfully identifies open-shell candidates 
even within the $\Delta N_{\text{Clar}}=0$ class, which the 
conventional picture would dismiss as non-radical. These 
BEM-predicted structures exhibit weak diradical character 
($y_{0}\approx 0.1$--$0.2$), consistent with the modest 
stabilization expected from resonance-space expansion alone. 
Particularly, we have verified the existence of unpaired electrons in 
these $\Delta N_{\text{Clar}}=0$ candidates using multireference 
wave-function methods (see SM), 
confirming that the BEM does not produce false positives but instead 
captures genuine, albeit weak, radical character that is invisible to 
Clar counting. These results establish the BEM as a robust and 
computationally efficient screening tool for identifying open-shell 
GNFs irrespective of whether their radical origin follows the 
Clar-sextet-gain or the resonance-space-expansion route.

To further elucidate the physical origin of unpaired electrons in 
Kekul\'{e} GNFs, we treat all structures as closed-shell and compute 
the BEM-predicted C--C bond occupancy number under the close-shell 
constraint. As shown in Figs.~\ref{fig3}(b) and \ref{fig3}(c), the 
bond ON distributions exhibit striking differences between 
genuine open-shell and closed-shell structures. For structures that 
are confirmed to possess open-shell character (e.g., structures 
\textbf{1}--\textbf{4} in Fig.~\ref{fig2}), the closed-shell BEM 
solution develops pronounced accumulations at occupancy number near 1, 
1.5, and 2, producing a trimodal distribution with sharp peaks. This 
signifies strong bond localization, in which the perfect-matching constraint 
forces a substantial fraction of C--C bonds into single-bond and 
double-bond limits, thereby suppressing the resonance manifold and 
creating an aromaticity deficit. In contrast, truly closed-shell 
structures (e.g., \textbf{5} and \textbf{6}) display a unimodal 
distribution centered near 1.5 with negligible weight near 2, 
indicating a naturally delocalized $\pi$ network that satisfies 
the maximum-entropy principle without frustration.

These observations reveal that the emergence of radicals in Kekul\'{e} 
GNFs is driven by an internal bond frustration, when the 
closed-shell pairing constraint compels too many bonds toward 
integral single- and double-bond limits, the number of accessible 
Clar resonators is severely restricted and the overall aromatic 
stabilization is insufficient. Breaking paired electrons, thereby 
allowing a finite $N_{u}$, relieves this constraint and enables the 
 ONs to redistribute away from the localized 1 and 2 
limits toward a more homogeneous and intermediate distribution. The 
resulting delocalization increases the configurational entropy of 
the bonding electrons and restores aromaticity across the entire 
$\pi$ network, providing the thermodynamic driving force for 
open-shell formation even in the absence of a net gain in Clar 
sextets.

To visualize the ON frustration mechanism concretely, we select two representative open-shell structures and compare their BEM-predicted C--C bond ON distributions under the closed-shell constraint and the optimized open-shell solution, as shown in Fig.~\ref{fig3}(c).  In the closed-shell constraint, the BEM yields a pronounced accumulation of ON near the double-bond limit ($n_{ij}\approx 2$). This produces a broad, high-intensity peak near occupancy 2 and a 
complementary peak near occupancy 1, reflecting the bond-length 
alternation characteristic of a frustrated, under-aromatized $\pi$ 
network. Upon releasing the closed-shell constraint and allowing a 
finite number of unpaired electrons, the bond-occupancy distribution 
becomes markedly more uniform, where the peak near occupancy 2 is 
substantially suppressed, and the ON redistribute toward 
intermediate values centered around 1.4--1.6. This delocalized 
distribution closely mirrors the C--C bond-length pattern obtained 
from DFT-B3LYP geometry optimizations, where the bond lengths 
converge to a narrower range between the single-bond and double-bond 
limits, confirming that the BEM captures the real-space structural 
relaxation accompanying the open-shell transition.

\begin{figure}
    \centering
    \includegraphics[width=\linewidth]{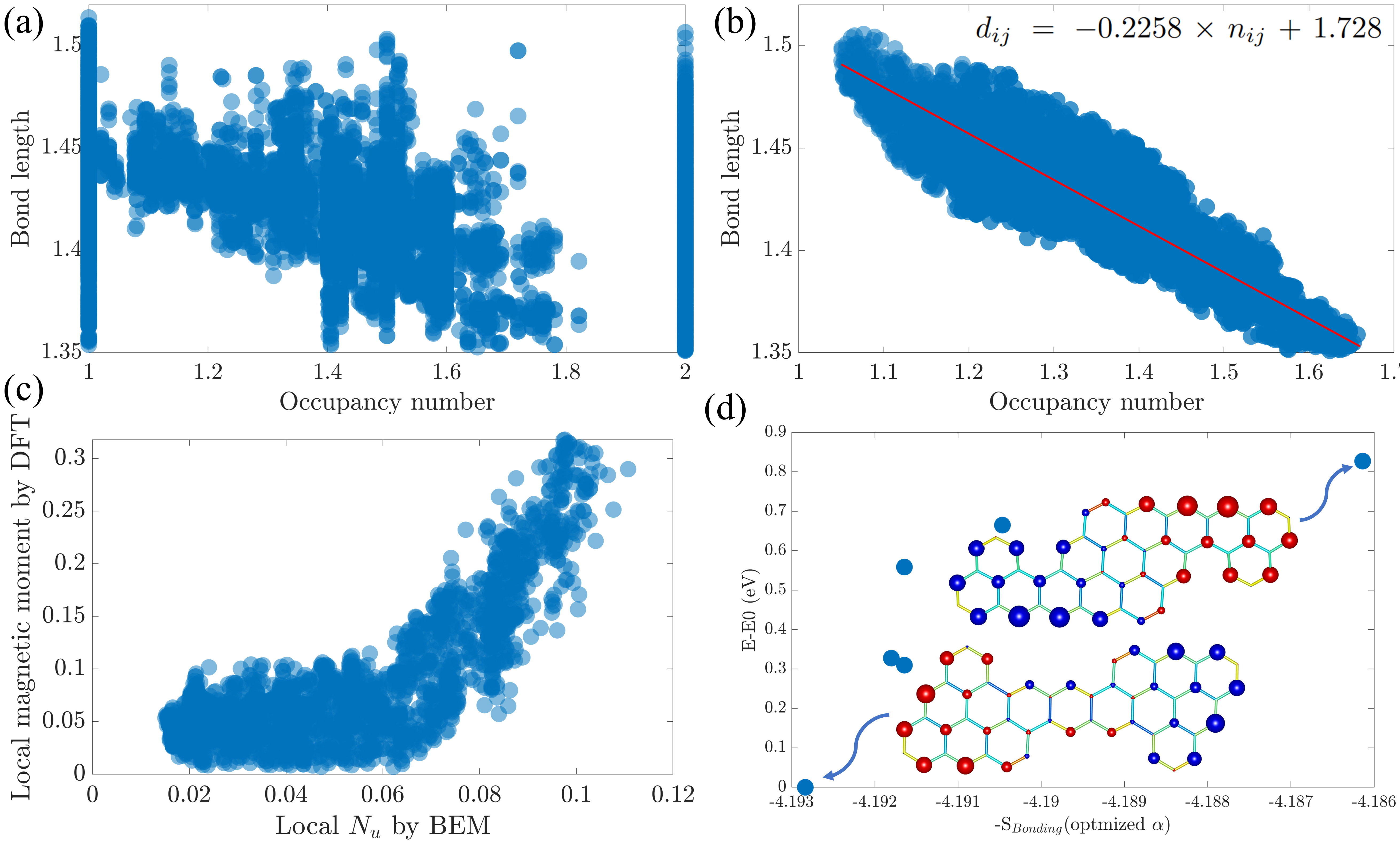}
\caption{\label{fig4} (a) BEM-predicted C--C bond-occupancy numbers 
versus DFT-optimized bond lengths under (a) the closed-shell  
and (b) the open-shell constraint for for diradical GNFs. (c) BEM versus DFT local magnetic 
moments. (d) Relative DFT energy versus BEM bonding entropy for 
\ce{C52H24} isomers.}
\end{figure}

To assess the predictive accuracy of the BEM for structural 
properties, we examine the correlation between BEM-predicted C--C 
bond ON and DFT-optimized bond lengths. Fig.~\ref{fig4}(a) 
shows the result obtained when open-shell Kekul\'{e} GNFs are 
forced into the closed-shell BEM constraint ($N_{u}=0$). A substantial 
fraction of the predicted C-C bond ON accumulate near the single-bond
($n_{ij}\approx 1$) and double-bond ($n_{ij}\approx 2$) limits, yet the 
corresponding C--C bond lengths scatter broadly from 1.35 to 1.52~\AA\ 
without any discernible correlation. In particular, bonds 
with ON near 1 or 2 span the entire observed length range, 
demonstrating that the closed-shell BEM completely fails to 
capture the bond-order--bond-length relationship in these 
radical-bearing structures.

For the enumerated Kekul\'{e} GNFs containing up to twelve hexagonal 
rings, the open-shell diradical character is consistently 
associated with two unpaired electrons ($N_{u}=2$). This allows us 
to determine the structure-specific weighting parameter $\alpha$ 
by enforcing $N_{u}=2$ in the BEM optimization. With this 
constraint, the open-shell BEM yields bond ON that 
exhibit a strong, linear correlation with the DFT-optimized C--C 
bond lengths across the entire dataset, as shown in 
Fig.~\ref{fig4}(b). Remarkably, the linear fit 
$d_{ij} = -0.2258 \times n_{ij}+1.728$ yields a universal slope and intercept that 
are essentially independent of the specific molecular topology, 
indicating that the BEM captures a general, transferable 
bond-order--bond-length relationship for Kekul\'{e} diradicals. 
This consistency confirms that the BEM, once calibrated by the 
correct unpaired-electron count, provides a quantitatively 
reliable and structurally predictive description of the 
electronic and geometric structure of open-shell GNFs.

For the diradical GNFs, the BEM predicts a site-resolved 
unpaired-electron population on each carbon atom that correlates 
positively with the DFT local magnetic moments, as shown in Fig.~\ref{fig4}(c). 
Notably, for BEM-predicted moments exceeding $0.06\,e$, the correlation with DFT is strongly 
linear, indicating high accuracy in the BEM description of robust spin polarization. Below this 
threshold, however, the BEM still predicts finite unpaired-electron densities where 
DFT yields nearly vanishing local moments. This discrepancy arises from the mean-field-like 
nature of the BEM, in which entropy maximization tends to homogenize the 
spin distribution across the $\pi$ network, whereas the B3LYP functional localizes 
magnetic moments onto specific carbon sites. The BEM therefore captures the 
global spin budget reliably but smooths over the spatial fragmentation inherent to DFT.

Fig.~\ref{fig4}(d) further validates the BEM by comparing its 
predicted bonding entropy with the relative DFT total energies for a 
series of \ce{C52H24} isomers with two unpaired electrons. The two quantities exhibit a strong 
monotonic correlation, confirming that the BEM captures the structural 
stability of Kekul\'{e} GNFs. The highest-energy isomer (upper right 
inset) features two pairs of mutually repelling H atoms in close 
proximity, in which the resulting steric and Coulomb repulsion forces the 
adjacent C--C bonds toward higher, more localized occupancies, thereby 
reducing the overall bonding entropy. Conversely, the lowest-energy 
isomer (lower left inset) maintains large H--H separations throughout, 
allowing the C--C bond occupancies to remain uniformly delocalized and 
the bonding entropy to reach its maximum. The BEM thus resolves subtle 
geometric frustrations through the bond-occupancy distribution, 
establishing bonding entropy as a reliable descriptor for the 
relative stability of open-shell carbon nanostructures.

\section{Conclusions}

We have proposed and computationally validated a second route to 
open-shell character in Kekul\'{e} graphene nanoflakes. In addition 
to the familiar gain of Clar aromatic sextets, a molecule can 
stabilize an open-shell state while retaining the same maximum 
Clar number if electron unpairing expands the number of 
equal-Clar resonance configurations. This Clar-number-invariant resonance-space expansion releases 
closed-shell pairing constraints, enables the migration of 
sextets through a larger configuration manifold, and produces a 
more delocalized distribution of C--C bond occupancies. The 
bonding entropy model provides a unified statistical framework 
that captures both the Clar-sextet-gain and resonance-space-expansion 
routes through an identical entropy-maximization principle, 
without requiring a priori knowledge of which mechanism 
is operative. Its optimized occupancies correlate with C--C bond 
lengths, local magnetic moments, DFT diradical character, and 
relative energies, while its predicted radical populations 
distinguish representative open- and closed-shell GNFs across a 
broad chemical space.  More broadly, the present work reframes Kekul\'{e} radical formation as a 
competition between electron-pairing cost and access to a larger 
bonding-configuration space, enabling graph-based screening of 
magnetic carbon nanostructures and offering a route to tune 
spin, exchange coupling, and optical response without requiring 
additional Clar sextets.

\begin{acknowledgments}
This work is supported by the Guangdong Basic and Applied Basic Research Foundation (Grants No. 2023A1515110894), the National Natural Science Foundation of China (Grant No. 12074126). This work is partially supported by High Performance Computing Platform of South China University of Technology. 
\end{acknowledgments}

\bibliography{ref}

@misc{github_bem,
  author = {He, C.-C.},
  title = {{BEM-for-kekule-GNFs}},
  howpublished = {\url{https://github.com/ChangChunHe/BEM-for-kekule-GNFs}},
  year = {2025},
  note = {Accessed: 2026-08-20}
}

@article{10.1021/acs.jpca.1c08661,
    author = {Wang, Yang},
    title = {Quantitative Resonance Theory Based on the Clar Sextet
Model},
    journal = {J. Phys. Chem. A},
    volume = {126},
    number = {1},
    pages = {164-176},
    year = {2022},
    month = {01},
    issn = {1089-5639},
    doi = {10.1021/acs.jpca.1c08661},
    url = {https://doi.org/10.1021/acs.jpca.1c08661},
}

@article{10.1021/jacs.2c11431,
  author = {Biswas, Kalyan and Soler, Diego and Mishra, Shantanu and others},
  title = {Steering Large Magnetic Exchange Coupling in Nanographenes near the Closed-Shell to Open-Shell Transition},
  journal = {J. Am. Chem. Soc.},
  volume = {145},
  number = {5},
  pages = {2968},
  year = {2023},
  month = {Jan},
  doi = {10.1021/jacs.2c11431}
}

@article{ditchfield1971self,
  title={Self-Consistent Molecular-Orbital Methods. {IX}. {An} Extended {Gaussian}-Type Basis for Molecular-Orbital Studies of Organic Molecules},
  author={Ditchfield, R. and Hehre, W. J. and Pople, J. A.},
  journal={J. Chem. Phys.},
  volume={54},
  pages={724},
  year={1971},
  doi={10.1063/1.1674902}
}

@article{hariharan1973influence,
  title={The Influence of Polarization Functions on Molecular Orbital Hydrogenation Energies},
  author={Hariharan, P. C. and Pople, J. A.},
  journal={Theor. Chim. Acta},
  volume={28},
  pages={213},
  year={1973},
  doi={10.1007/BF00533485}
}

@article{10.1063/5.0280786,
    author = {Luo, Hai-Wei and He, Chang-Chun and Zhao, Yu-Jun and Yang, Xiao-Bao},
    title = {Bandgap opening induced by electron localization in graphene antidot lattices},
    journal = {J. Chem. Phys. },
    volume = {163},
    number = {7},
    pages = {074701},
    year = {2025},
    month = {08},
    issn = {0021-9606},
    doi = {10.1063/5.0280786},
    url = {https://doi.org/10.1063/5.0280786},
}

@misc{g16,
  title = {Gaussian~16, {R}evision~C.01},
  author = {Frisch, M. J. and Trucks, G. W. and Schlegel, H. B. and Scuseria, G. E. and Robb, M. A. and Cheeseman, J. R. and Scalmani, G. and Barone, V. and Petersson, G. A. and Nakatsuji, H. and Li, X. and Caricato, M. and Marenich, A. V. and Bloino, J. and Janesko, B. G. and Gomperts, R. and Mennucci, B. and Hratchian, H. P. and Ortiz, J. V. and Izmaylov, A. F. and Sonnenberg, J. L. and Williams-Young, D. and Ding, F. and Lipparini, F. and Egidi, F. and Goings, J. and Peng, B. and Petrone, A. and Henderson, T. and Ranasinghe, D. and Zakrzewski, V. G. and Gao, J. and Rega, N. and Zheng, G. and Liang, W. and Hada, M. and Ehara, M. and Toyota, K. and Fukuda, R. and Hasegawa, J. and Ishida, M. and Nakajima, T. and Honda, Y. and Kitao, O. and Nakai, H. and Vreven, T. and Throssell, K. and Montgomery, J. A., Jr. and Peralta, J. E. and Ogliaro, F. and Bearpark, M. J. and Heyd, J. J. and Brothers, E. N. and Kudin, K. N. and Staroverov, V. N. and Keith, T. A. and Kobayashi, R. and Normand, J. and Raghavachari, K. and Rendell, A. P. and Burant, J. C. and Iyengar, S. S. and Tomasi, J. and Cossi, M. and Millam, J. M. and Klene, M. and Adamo, C. and Cammi, R. and Ochterski, J. W. and Martin, R. L. and Morokuma, K. and Farkas, O. and Foresman, J. B. and Fox, D. J.},
  year = {2016},
  note = {Gaussian, Inc., Wallingford CT}
}

@article{PhysRevB.111.085408,
  title = {Entropy-driven electron density and effective model Hamiltonian for boron systems},
  author = {He, Chang-Chun and Xu, Shao-Gang and Zhao, Yu-Jun and Xu, Hu and Yang, Xiao-Bao},
  journal = {Phys. Rev. B},
  volume = {111},
  issue = {8},
  pages = {085408},
  numpages = {9},
  year = {2025},
  month = {Feb},
  publisher = {American Physical Society},
  doi = {10.1103/PhysRevB.111.085408},
  url = {https://link.aps.org/doi/10.1103/PhysRevB.111.085408}
}

@article{kuriakose2022design,
  title={Design and Synthesis of {Kekul\'{e}} and Non-{Kekul\'{e}} Diradicaloids via the Radical Periannulation Strategy: The Power of Seven {Clar's} Sextets},
  author={Kuriakose, F. and Commodore, M. and Hu, C. and Fabiano, C. J. and Sen, D. and Li, R. R. and Bisht, S. and {\"U}ng{\"o}r, {\"O}. and Lin, X. and Strouse, G. F. and DePrince, A. E. I. and Lazenby, R. A. and Mentink-Vigier, F. and Shatruk, M. and Alabugin, I. V.},
  journal={J. Am. Chem. Soc.},
  volume={144},
  pages={23448},
  year={2022},
  doi={10.1021/jacs.2c09637}
}

@article{shu2023stable,
  title={From stable radicals to thermally robust high-spin diradicals and triradicals},
  author={Shu, C. and Yang, Z. and Rajca, A.},
  journal={Chem. Rev.},
  volume={123},
  pages={11954},
  year={2023},
  doi={10.1021/acs.chemrev.3c00406}
}

@article{pan2024magnetooptical,
  title={Magnetooptical studies of organic electroluminescent materials having fast reverse intersystem crossing},
  author={Pan, X. and Khanal, D. R. and Kwon, O. and Vardeny, Z. V.},
  journal={Phys. Rev. Appl.},
  volume={21},
  pages={034057},
  year={2024},
  doi={10.1103/PhysRevApplied.21.034057}
}

@article{weng2024biolympicenyl,
  title={1,1'-Biolympicenyl: A stable non-{Kekul\'{e}} diradical with a small singlet and triplet energy gap},
  author={Weng, T. and Xu, Z. and Li, K. and Guo, Y. and Chen, X. and Li, Z. and Sun, Z.},
  journal={J. Am. Chem. Soc.},
  volume={146},
  pages={26454},
  year={2024},
  doi={10.1021/jacs.4c09627}
}

@article{pavlicek2017synthesis,
  title={Synthesis and characterization of triangulene},
  author={Pavli{\v{c}}ek, N. and Mistry, A. and Majzik, Z. and Moll, N. and Meyer, G. and Fox, D. J. and Gross, L.},
  journal={Nat. Nanotechnol.},
  volume={12},
  pages={308},
  year={2017},
  doi={10.1038/nnano.2016.305}
}

@article{mishra2020topological,
  title={Topological frustration induces unconventional magnetism in a nanographene},
  author={Mishra, S. and Beyer, D. and Eimre, K. and Kezilebieke, S. and Berger, R. and Gr{\"o}ning, O. and Pignedoli, C. A. and M{\"u}llen, K. and Liljeroth, P. and Ruffieux, P. and Feng, X. and Fasel, R.},
  journal={Nat. Nanotechnol.},
  volume={15},
  pages={22},
  year={2020},
  doi={10.1038/s41565-019-0577-9}
}

@book{clar1972aromatic,
  title={The Aromatic Sextet},
  author={Clar, E.},
  publisher={Wiley},
  address={London},
  year={1972}
}

@article{bendikov2004oligoacenes,
  title={Oligoacenes: Theoretical prediction of open-shell singlet diradical ground states},
  author={Bendikov, M. and Duong, H. M. and Starkey, K. and Houk, K. N. and Carter, E. A. and Wudl, F.},
  journal={J. Am. Chem. Soc.},
  volume={126},
  pages={7416},
  year={2004},
  doi={10.1021/ja048919w}
}

@article{trinquier2018predicting,
  title={Predicting the open-shell character of polycyclic hydrocarbons in terms of {Clar} sextets},
  author={Trinquier, G. and Malrieu, J.-P.},
  journal={J. Phys. Chem. A},
  volume={122},
  pages={1088},
  year={2018},
  doi={10.1021/acs.jpca.7b11095}
}

@article{yeh2016role,
  title={Role of {Kekul\'{e}} and non-{Kekul\'{e}} structures in the radical character of alternant polycyclic aromatic hydrocarbons: A {TAO-DFT} study},
  author={Yeh, C.-N. and Chai, J.-D.},
  journal={Sci. Rep.},
  volume={6},
  pages={30562},
  year={2016},
  doi={10.1038/srep30562}
}

@article{gopalakrishna2018open,
  title={From open-shell singlet diradicaloids to polyradicaloids},
  author={Gopalakrishna, T. Y. and Zeng, W. and Lu, X. and Wu, J.},
  journal={Chem. Commun.},
  volume={54},
  pages={2186},
  year={2018},
  doi={10.1039/C7CC09949E}
}

@article{liu2015tetrabenzo,
  title={Tetrabenzo[a,f,j,o]perylene: A polycyclic aromatic hydrocarbon with an open-shell singlet biradical ground state},
  author={Liu, J. and Ravat, P. and Wagner, M. and Baumgarten, M. and Feng, X. and M{\"u}llen, K.},
  journal={Angew. Chem. Int. Ed.},
  volume={54},
  pages={12442},
  year={2015},
  doi={10.1002/anie.201505245}
}

@article{noda2018excited,
  title={Excited state engineering for efficient reverse intersystem crossing},
  author={Noda, H. and Nakanotani, H. and Adachi, C.},
  journal={Sci. Adv.},
  volume={4},
  pages={eaao6910},
  year={2018},
  doi={10.1126/sciadv.aao6910}
}

@article{hosoya1981graph,
  title={Graph-theoretical analysis of {Clar's} aromatic sextet: Mathematical properties of the set of the {Kekul\'{e}} patterns and the sextet polynomial for polycyclic aromatic hydrocarbons},
  author={Hosoya, H. and Hosoi, K. and Gutman, I.},
  journal={Tetrahedron},
  volume={37},
  pages={1113},
  year={1981},
  doi={10.1016/S0040-4020(01)92040-X}
}

@article{he2025graph,
  title={A graph-based statistical model for carbon nanostructures},
  author={He, C.-C. and Xu, S.-G. and Zeng, J. and Huang, W. and Yao, Y. and Zhao, Y.-J. and Xu, H. and Yang, X.-B.},
  journal={J. Chem. Phys.},
  volume={162},
  pages={154104},
  year={2025},
  doi={10.1063/5.0244219}
}

@article{he2025unified,
  title={Unified bonding entropy model to determine magnetic properties in graphene nanoflakes},
  author={He, C.-C. and Zeng, J. and Zhao, Y.-J. and Yang, X.-B.},
  journal={Phys. Rev. B},
  volume={112},
  pages={094404},
  year={2025},
  doi={10.1103/PhysRevB.112.094404}
}

\end{document}